\documentclass[modern]{aastex63}

\begin{document}

\title{Simultaneous Six-way Observations from the Navy Precision Optical Interferometer}

\author{Ellyn K. Baines}
\affil{Naval Research Laboratory, Remote Sensing Division, 4555 Overlook Ave SW, Washington, DC 20375, USA}
\email{ellyn.k.baines.civ@us.navy.mil}

\author{Solvay Blomquist}
\affil{Lowell Observatory, 1400 W. Mars Hill Rd, Flagstaff, AZ 86001, USA} 

\author{James H. Clark III}
\affil{Naval Research Laboratory, Remote Sensing Division, 4555 Overlook Ave SW, Washington, DC 20375, USA}

\author{Jim Gorney}
\affil{Lowell Observatory, 1400 W. Mars Hill Rd, Flagstaff, AZ 86001, USA}

\author{Erin Maier}
\affil{Lowell Observatory, 1400 W. Mars Hill Rd, Flagstaff, AZ 86001, USA} 

\author{Jason Sanborn}
\affil{Lowell Observatory, 1400 W. Mars Hill Rd, Flagstaff, AZ 86001, USA} 

\author{Henrique R. Schmitt}
\affil{Naval Research Laboratory, Remote Sensing Division, 4555 Overlook Ave SW, Washington, DC 20375, USA}

\author{Jordan M. Stone}
\affil{Naval Research Laboratory, Remote Sensing Division, 4555 Overlook Ave SW, Washington, DC 20375, USA}

\author{Gerard T. van Belle}
\affil{Lowell Observatory, 1400 W. Mars Hill Rd, Flagstaff, AZ 86001, USA} 

\author{Kaspar von Braun}
\affil{Lowell Observatory, 1400 W. Mars Hill Rd, Flagstaff, AZ 86001, USA}

\begin{abstract}

We measured the angular diameters of six stars using the 6-element observing mode of the Navy Precision Optical Interferometer (NPOI) for the first time since the early 2000s. Four of the diameters ranged from 1.2 mas to 1.9 mas, while the two others were much smaller at approximately 0.5 mas to 0.7 mas, which are the two smallest angular diameters measured to date with the NPOI. There is a larger spread in the measurements than data obtained with 3- or 4- or 5-element modes, which can be attributed in part to the flux imbalance due to the combination of more than 2 siderostats in a single spectrograph, and also to cross talk between multiple baselines related to non-linearities in the fast delay line dither strokes. We plan to address this in the future by using the VISION beam combiner.

\end{abstract}



\section{Introduction} 

The Navy Precision Optical Interferometer (NPOI) has been in operation since 1994, originally with the name Navy Prototype Optical Interferometer and then, briefly, Navy Optical Interferometer. It is a Y-shaped optical interferometer located on Anderson Mesa near Flagstaff, AZ. The NPOI was originally designed to combine light from six elements\footnote{The NPOI observes with siderostats, and in this paper ``6-way observing'' means we observed using 6 siderostats simultaneously.} as a balance between financial cost, instrument complexity, and the tension between a philosophy of ``if some is good, more is better'' and the dilution of fringes across multiple apertures \citep{1998ApJ...496..550A}. 

The NPOI consists of two nested subarrays that can be combined at will, depending on the requirements for specific scientific questions. The four stations of the astrometric array are fixed near the center of the Y, and are named AC, AE, AN, and AW, which stand for astrometric center, east, north, and west, respectively. The other subarray is the imaging array spread along the three north-, east-, and west-oriented arms of the NPOI. Each arm has ten piers on which a siderostat can be placed, meaning the imaging array can be reconfigured as needed. The stations are labeled according to which arm they are on and how far away they are from the array center, with 1 being closest and 10 being farthest away. This paper includes data from AC, AE, and AW of the astrometric array, and E6, W4, and W7 of the imaging array. 

The NPOI has used many different configurations through the years, from a single baseline (i.e., two imaging elements where the ``baseline'' is the distance between them) up to 6-way beam combination. Most of the time, the NPOI has used three to five siderostats at a time, which has the advantage of increasing sky coverage and the length of time a given target is observable as it moves through the available swath of sky. The NPOI first went on-sky in 6-way mode in September 2001 during on-sky tests, and routine observations began January 2002 \citep{2003SPIE.4838..358B}. Results from that time include imaging the triple star $\eta$ Virginis, modeling its orbit and detecting the motion of the close pair over time \citep{2003AJ....125.2630H}. Six-way mode was halted not long afterwards when other observational programs took precedence, and chronic problems with delay lines made this type of operation impractical.

One of the main issues that led to halting 6-way data collection at that time was the reduced sky coverage that can be achieved in 6-way when one uses the longest baselines without the long delay lines, usually of the order of 1 hour or less over a narrow range of declinations. The other more severe issue was the irregularities in the fast delay line (FDL) strokes and their truncated range of motion for the largest stroke amplitude of 4 $\mu$m \citep{2006SPIE.6268E..4AJ}, which resulted in cross talk between baselines with adjacent stroke frequencies. In addition to these issues, one of the delay lines was taken offline for an extended period of time in order to develop new FDL controllers. Although we were not able to address issues related limited sky coverage and stroke irregularities, we avoided the truncated range of motion issue by using only stroke amplitudes up to $\pm3$ $\mu$m.

This paper is organized as follows: Section 2 outlines our observing and data reduction procedures. Section 3 describes how we determine various stellar parameters such as angular diameters, physical radii, effective temperatures, bolometric flux, and luminosity. Section 4 presents notes on individual fits, when applicable, and plans for future 6-way observations.

\clearpage


\section{Interferometric Observations}

We observed six stars in 6-way mode in August and September of 2021, collecting nearly 23,000 calibrated data points. The stars were selected to be small ($\leq$2.0 mas) and bright ($V \leq$ 4.3) so that finding interferometric fringes on all tracking baselines would not present an undue challenge. Table \ref{general_properties} lists each star's identifiers, spectral type, $V$ magnitude, parallax, and metallicity. Table \ref{observations} is the observing log and includes the stars observed, their calibrators, dates, the baselines used, and number of data points per night (note that one of the stars also includes 5-way data taken in June 2021). We used the ``Classic'' beam combiner \citep{2003AJ....125.2630H, 2003SPIE.4838..358B, 2016ApJS..227....4H} that takes data across 16 spectral channels in the visible regime from 550 nm to 850 nm. 

\begin{figure}[h]
\includegraphics[width=1.0\textwidth]{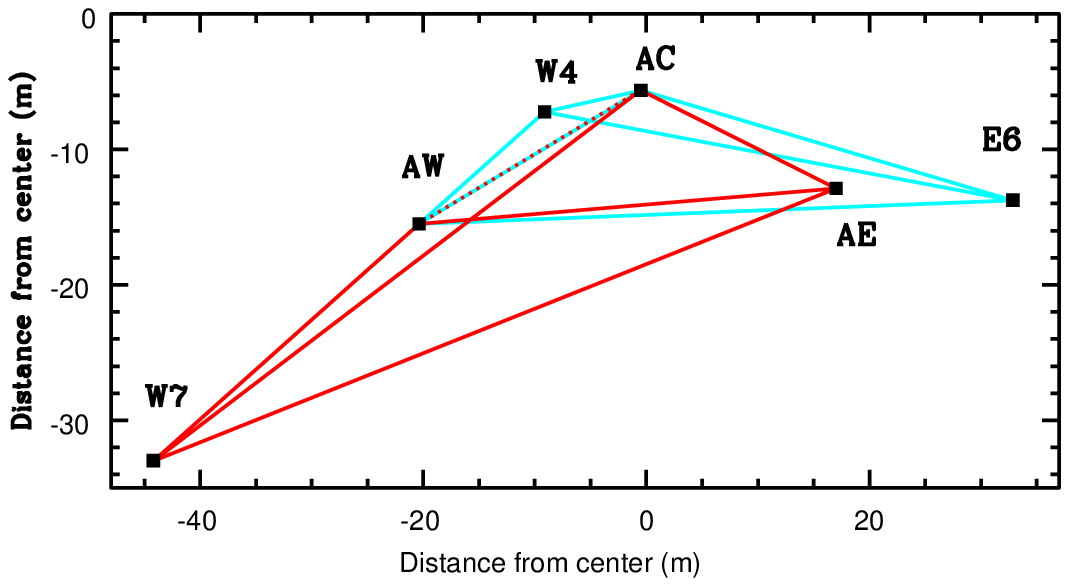}
\caption{The NPOI configuration used for 6-way observing. The squares show the locations of the siderostats as a function of distance from the center, the red lines show the baselines on spectrograph 1, and the blue lines show the baselines on spectrograph 2. The dashed line is the baseline that repeats on both spectrographs. Table \ref{baselines} lists the lengths of the various baselines.}
  \label{array_config}
\end{figure}

Hardware limitations prevent us from recording all 15 baselines possible with the 6 imaging elements, so these types of observations produce fringes on 11 simultaneous baselines. This is because we use two spectrographs with four siderostats on each, giving us six baselines per spectrograph. One of those baselines is repeated on each spectrograph, which is how we end up with 11 baselines per observation. Table \ref{baselines} shows a list of the baselines used, and Figure \ref{array_config} shows the configuration.

We interleaved scans on the target stars with scans of calibrator stars to help minimize errors introduced by atmospheric turbulence and instrumental imperfections. We chose calibrator stars with small angular diameters\footnote{Here, ``small'' means that the star's diameter is significantly less than the resolution of the interferometer.} and checked for binarity, variability, and rapid rotation. Some of the calibrator stars used featured one or more of those characteristics, but not to an extent that would affect the calibration process: any binary separations or brightness ratios were beyond the detection limit of the configuration used, while oblateness due to rapid rotation and/or variability did not introduce a variation in the diameter of the star that would be large enough to cause significant calibration issues.

To estimate the calibrator stars' angular diameters, we created spectral energy distribution (SED) fits based on published $UBVRIJHK$ photometric values. We used plane-parallel model atmospheres \citep{2003IAUS..210P.A20C} based on effective temperature ($T_{\rm eff}$), surface gravity (log~$g$), and $E(B-V)$. Stellar models were fit to observed photometry after converting the magnitudes to fluxes using \citet{1996AJ....112..307C} for $UBVRI$ and \citet{2003AJ....126.1090C} for $JHK$. Table \ref{calibrators} lists the photometry, $T_{\rm eff}$, log~$g$, and $E(B-V)$ used, and the resulting angular diameters. This is a simple SED fit, unlike the more sophisticated one described in Section 3.2 that we used for the target stars. It is an appropriate method for calibrator stars, given the insensitivity of the target's measured angular diameter with respect to the calibrator's diameter \citep{2018AJ....155...30B}.

Each observation consisted of a 30-second coherent (on the fringe) scan where the fringe contrast was measured every 2 ms. Every coherent scan was paired with an incoherent (off the fringe) scan, which acted as an estimate for the additive bias affecting fringe measurements \citep{2003AJ....125.2630H}. Each coherent scan was averaged to 1-second data points, and then again to a single 30-second average. The dispersion of the 1-second data points served as an estimate of the internal uncertainties. The NPOI's data reduction package $OYSTER$ was developed by C. A. Hummel\footnote{www.eso.org/$\sim$chummel/oyster/oyster.html} and automatically edits data using the method described in \citet{2003AJ....125.2630H}.

In addition to the automated process, we edited out individual data points and/or scans that showed large scatter, on the order of 5-$\sigma$ or higher. This was more common in the channels corresponding to shorter wavelengths where the spectral channels are narrower, atmospheric effects are more pronounced, and the avalanche photodiode detectors have lower quantum efficiency. Removing these points did not affect the diameter measurements. 


\section{Determining Stellar Parameters}

\subsection{Angular Diameter Measurement}

Interferometric diameter measurements use visibility squared ($V^2$). For a point source, $V^2$ is 1 and it is considered completely unresolved, while a star is defined as completely resolved when its $V^2$ reaches zero. For a uniformly-illuminated disk, $V^2 = [2 J_1(x) / x]^2$, where $J_1$ is the Bessel function of the first order, $x = \pi B \theta_{\rm UD} \lambda^{-1}$, $B$ is the projected baseline toward the star's position, $\theta_{\rm UD}$ is the apparent uniform disk angular diameter of the star, and $\lambda$ is the effective wavelength of the observation \citep{1992ARAandA..30..457S}. $\theta_{\rm UD}$ results for our program stars are listed in Table \ref{inf_results}, and the data are freely available in OIFITS form \citep{2017AandA...597A...8D} upon request.

A more realistic description of a star's surface brightness includes limb darkening (LD).  If a linear LD coefficient $\mu_\lambda$ is used, then
\begin{eqnarray}
V^2 = \left( {1-\mu_\lambda \over 2} + {\mu_\lambda \over 3} \right)^{-1} 
\times 
\left[(1-\mu_\lambda) {J_1(x_{\rm LD}) \over x_{\rm LD}} + \mu_\lambda {\left( \frac{\pi}{2} \right)^{1/2} \frac{J_{3/2}(x_{\rm LD})}{x_{\rm LD}^{3/2}}} \right]^2 .
\end{eqnarray}
where $x_{\rm LD} = \pi B\theta_{\rm LD}\lambda^{-1}$ \citep{1974MNRAS.167..475H}. We used $T_{\rm eff}$, log $g$, and metallicity ([Fe/H]) values from the literature with an assumed microturbulent velocity of 2 km s$^{\rm -1}$ to obtain $\mu_\lambda$ from \citet{2011AandA...529A..75C}. We used the ATLAS stellar model in the \emph{R}-band, the waveband most closely matched to the central wavelength of the NPOI's bandpass. A more sophisticated analysis of these stars would include the non-linear nature of limb darkening, and how it depends on wavelength. The simpler treatment here is valid, given that the strength of the limb darkening for star is related to the height of the second maximum of the visibility curve \citep{2001AandA...377..981W}, and none of our measurements were beyond the first minimum.

\begin{figure}[h]
\includegraphics[width=1.0\textwidth]{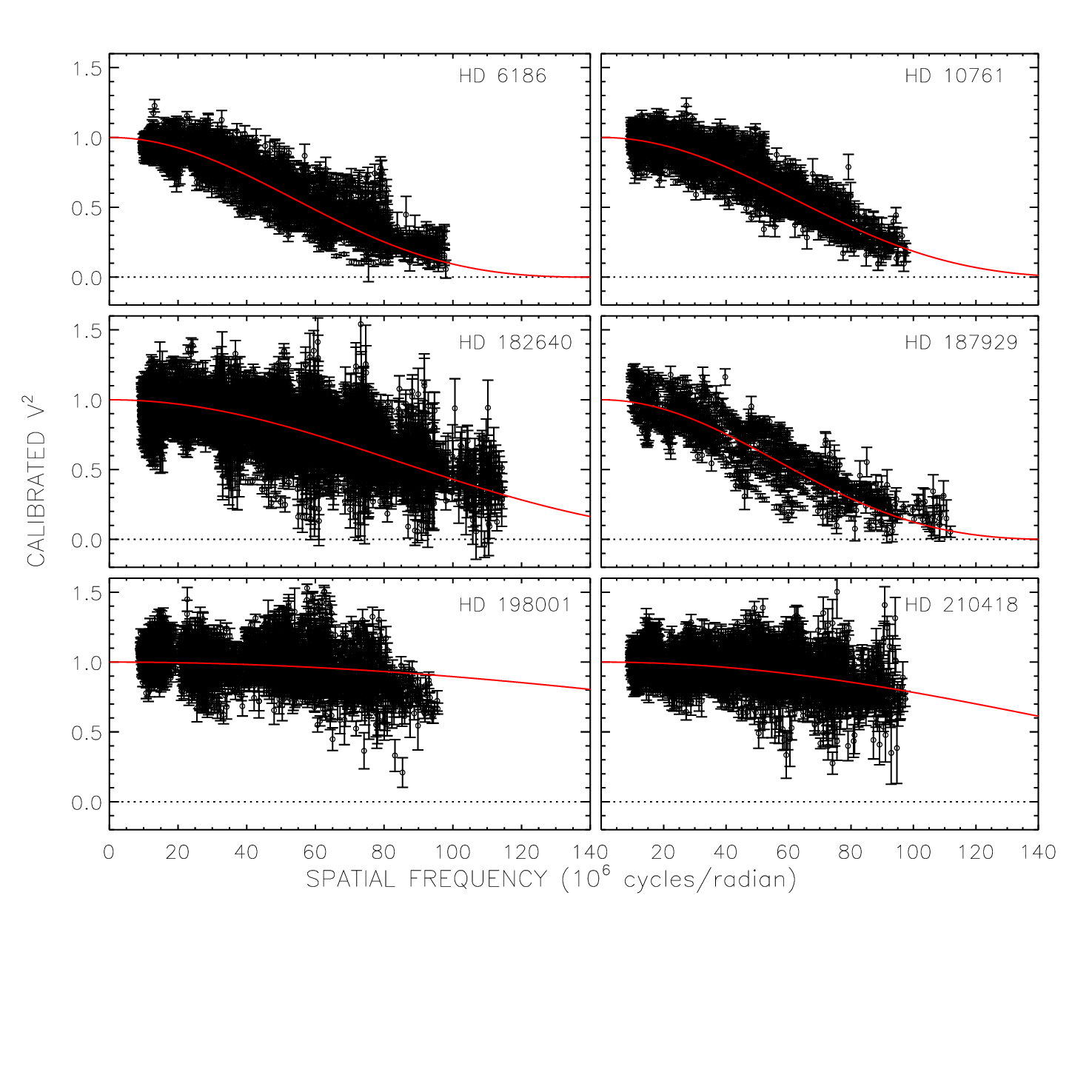}
\caption{Angular diameter fits to measured visibilities. The solid red line represents the visibility curve for the best fit $\theta_{\rm LD}$, the open circles are the calibrated visibilities, and the vertical lines are the measurement uncertainties. }
  \label{vis_plots}
\end{figure}

The $T_{\rm eff}$, log $g$, and $\mu_\lambda$ used and the resulting limb darkened diameters ($\theta_{\rm LD}$) are listed in Table \ref{inf_results} along with the maximum spatial frequency for each star's data set, and the number of data points in the angular diameter fit. Figure \ref{vis_plots} shows the $\theta_{\rm LD}$ fits for the six stars. 

\begin{figure}[h]
\includegraphics[width=1.0\textwidth]{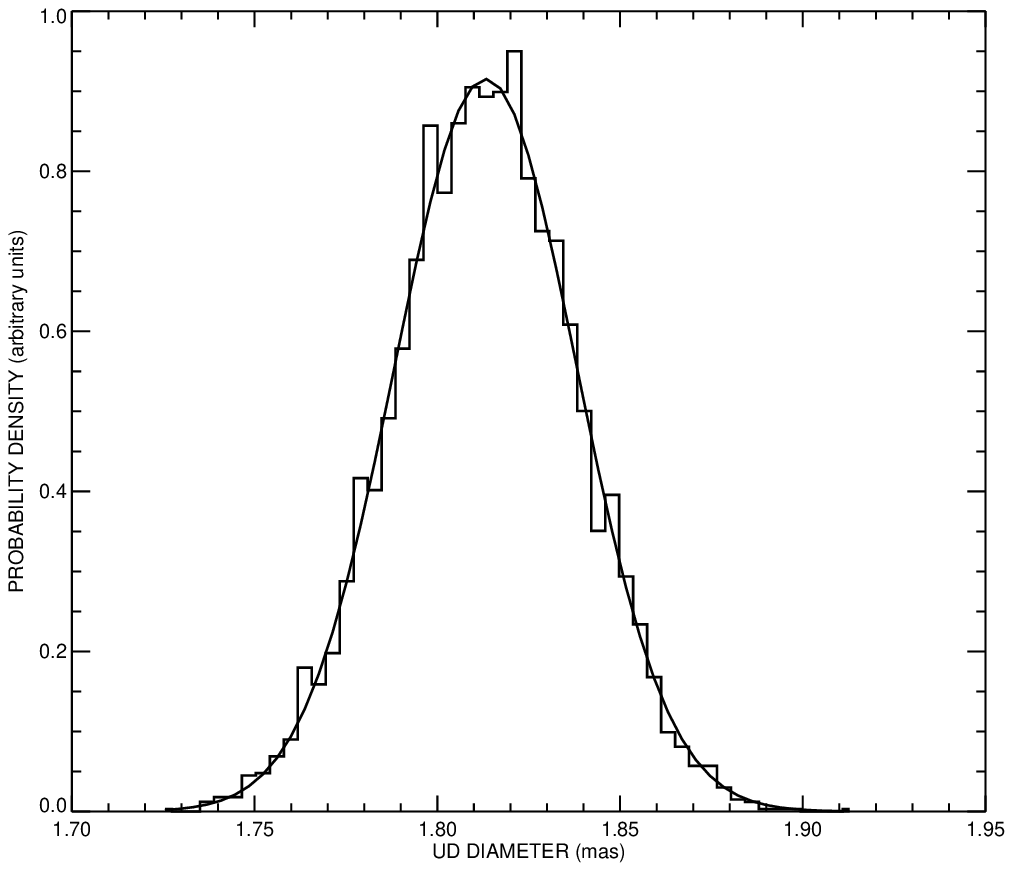}
\caption{An example probability density solution for the diameter fit to HD 6168/$\epsilon$ Psc visibilities as described in Section 3.1.}
  \label{plot_gauss}
\end{figure}

We used the procedure described in \citet{2018AJ....155...30B} to estimate angular diameter uncertainties, which can be summarized thus: if we fit only the collected data points without regard to correlations within a scan, the diameter's uncertainty can be significantly underestimated. To address this, we used a modified bootstrap Monte Carlo method developed by \citet{2010SPIE.7734E.103T} to generate a large number of synthetic data sets by randomly selecting entire scans. The width of the distribution of diameters fit to these data sets becomes our measure of the uncertainty for the diameter (see Figure \ref{plot_gauss}).


\subsection{Stellar Radius, Luminosity, and Effective Temperature}

Our next step was to convert our angular diameters to stellar sizes in solar radii. When available, the parallax from the Gaia Data Release 3 \citep{Gaia2022} was converted into a distance and combined with our measured diameters to calculate the physical radius $R$. Otherwise, parallaxes from \citet{2007AandA...474..653V} and \citet{2018AandA...616A...1G} were used. 

In order to determine each star's luminosity ($L$) and $T_{\rm eff}$, we generated SED fits using photometric values published in \citet{1966CoLPL...4...99J}, \citet{1972VA.....14...13G}, \citet{1975RMxAA...1..299J}, \citet{1984AandAS...57..357O}, \citet{1987AandAS...71..413M}, \citet{1988iras....7.....H}, \citet{1988iras....1.....B}, \citet{1989AandAS...81..401M}, \citet{Mermilliod}, \citet{1993cio..book.....G}, \citet{1993AandAS..100..591O}, \citet{1999yCat.2225....0G}, \citet{2000AandA...355L..27H}, \citet{2002yCat.2237....0D}, \citet{2003yCat.2246....0C}, \citet{2004ApJS..154..673S}, and \citet{2007AandA...474..653V}. The assigned uncertainties for the 2MASS infrared measurements are as reported in \citet{2003yCat.2246....0C}, and an uncertainty of 0.05 mag was assigned to the optical measurements. 

Spectrophotometry from \citet{1996yCat.3126....0B}, \citet{1983TrSht..53...50G}, and \citet{1988scs..book.....K} were included for HD 6186/$\epsilon$ Psc, HD 10761/o Psc, HD 182640/$\delta$ Aql, but not HD 198001/$\epsilon$ Aqr and HD 210418/$\theta$ Peg. HD 187929/$\eta$ Aql is a well-known Cepheid variable, and the SED fit did not produce usable results so the remainder of the following calculations apply to the remaining five stars.

We determined the best fit stellar spectral template to the photometry and spectrophotometry, if used, from the flux-calibrated stellar spectral atlas of \citet{1998PASP..110..863P} using the $\chi^2$ minimization technique \citep{1992nrca.book.....P, 2003psa..book.....W}. This provided the bolometric flux ($F_{\rm BOL}$) for each star and allowed for the calculation of extinction ($A_{\rm V}$) with the wavelength-dependent reddening relations of \citet{1989ApJ...345..245C}.

We combined our $F_{\rm BOL}$ values with the stars' distances to estimate $L$ using $L = 4 \pi d^2 F_{\rm BOL}$. We also combined the $F_{\rm BOL}$ with $\theta_{\rm LD}$ to determine each star's $T_{\rm eff}$ using the relation,
\begin{equation}
F_{\rm BOL} = {1 \over 4} \theta_{\rm LD}^2 \sigma T_{\rm eff}^4,
\end{equation}
where $\sigma$ is the Stefan-Boltzmann constant and $\theta_{\rm LD}$ is in radians \citep{1999AJ....117..521V}. The resulting $R$, $F_{\rm BOL}$, $A_{\rm V}$, $T_{\rm eff}$, and $L$ are listed in Table \ref{derived_results}.

Considering that $\mu_\lambda$ is chosen based on a given $T_{\rm eff}$, we used an iterative process to determine the final $\theta_{\rm LD}$. We began with the initial $\theta_{\rm LD}$ determined using the process described in Section 3.1, calculated $T_{\rm eff}$, and used that new $T_{\rm eff}$ to see if $\mu_\lambda$ was altered. The largest change in $\mu_\lambda$ for all the stars was 0.03, which made at most a 0.3$\%$ difference in $\theta_{\rm LD}$ (0.004 mas), well within the uncertainty on the diameter. Similarly, $T_{\rm eff}$ changed by a maximum of 11 K as $\mu_\lambda$ was updated. This procedure took one iteration for all the stars to get to the final $\theta_{\rm LD}$, $\mu_\lambda$, and $T_{\rm eff}$. The initial and final values for all three quantities are listed in Table \ref{inf_results}.

\section{Discussion and Conclusions}

For five of the six stars, the diameter fits are excellent and cover the majority of the visibility curve. The exception is HD 198001/$\epsilon$ Aqr. It is the smallest star ever measured with the NPOI at 0.503 mas, and the uncertainty of 0.357 mas is a sizable percentage of that diameter. Still, we find the measurement of value, even as we hope to improve on the uncertainty with future observations. 

Two of the stars have been previously measured using interferometry in the last 10 years: \citet{2021ApJ...922..163V} determined a diameter of 1.923$\pm$0.045 mas for HD 6186/$\epsilon$ Psc, compared to our measurement of 1.887$\pm$0.025 mas, and \citet{2012ApJ...746..101B} found a diameter of 0.862$\pm$0.018 mas for HD 210418/$\theta$ Peg, versus our 0.688$\pm$0.031 mas. Considering this is one of the smallest diameters ever measured with the NPOI and is below the resolution limit, this discrepancy is not surprising.

\begin{figure}[h]
\includegraphics[width=1.0\textwidth]{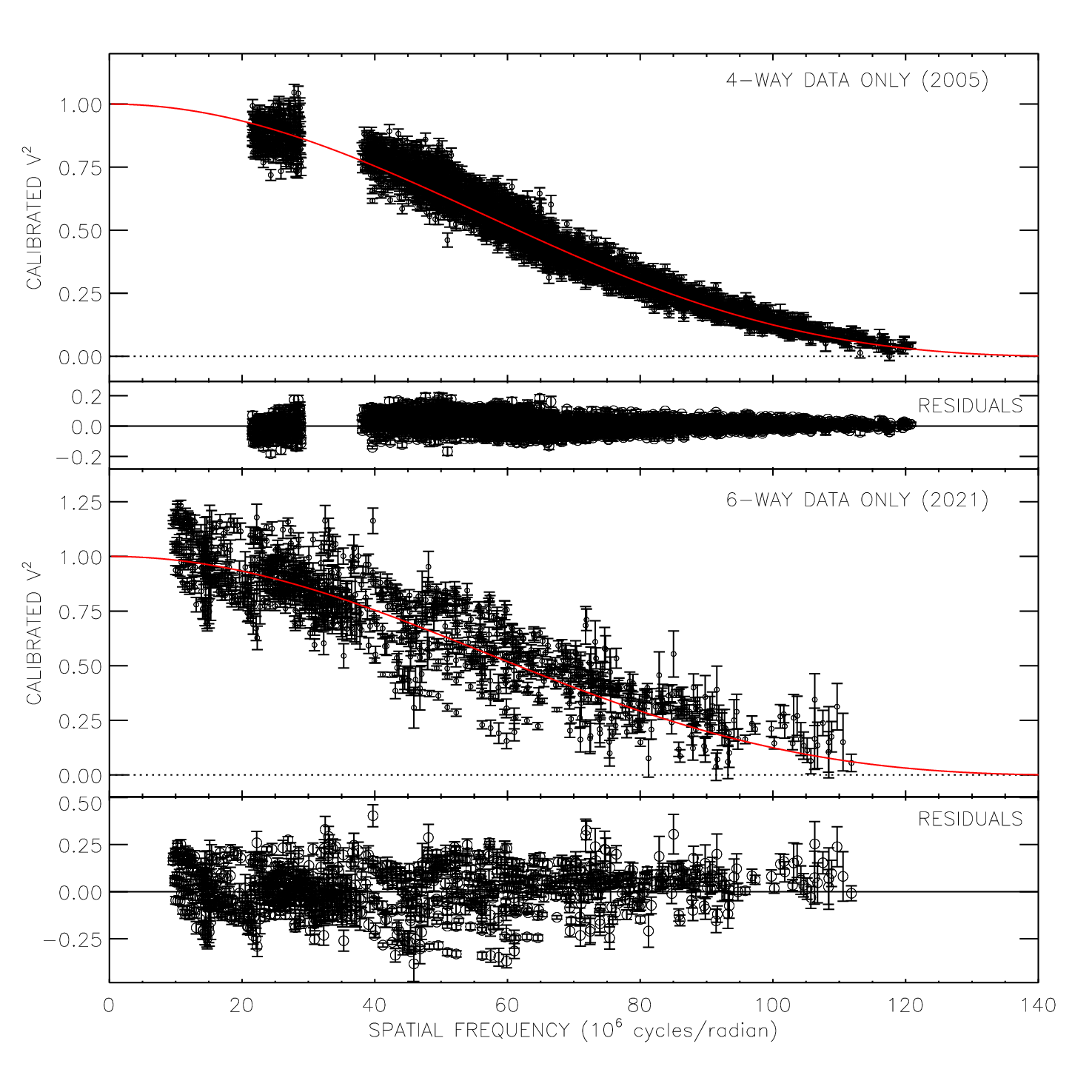}
\caption{Comparing the angular diameter fit using 4-way data (top half) from 2005 with that of 6-way data from 2021 (bottom half) for HD 187929/$\eta$ Aql. The symbols are the same as in Figure \ref{vis_plots}, while bottom portion of each half shows the residuals to the angular diameter fit. The 4-way data had 3 siderostats per spectrograph, while the 6-way data had 4 siderostats per spectrograph, which partially accounts for the increased scatter in the residuals for the latter.}
  \label{HD187929_plots}
\end{figure}

Interestingly, HD 187929/$\eta$ Aql was observed using 4-way data collection in 2005 (with 3 siderostats per spectrograph), and Figure \ref{HD187929_plots} shows how the older data compare to the 6-way data (with 4 siderostats per spectrograph). The diameter determined from the 4-way data is 1.804$\pm$0.007 mas \citep{2018AJ....155...30B}, and the diameter from the 6-way data is 1.808$\pm$0.055 mas. The 4-way data show a tighter fit to the visibility curve while the 6-way data have more spread around the best fit angular diameter. 

The larger spread in the visibilities and residuals for the 6-way data can be attributed to two effects: the reduction of the visibility amplitudes due to flux imbalance, and cross talk between the different baselines due to non linearities in the fast delay line modulation strokes \citep{2008poii.conf..571S}. In the case of flux imbalance, the $V^2$ of a baseline observed in a detector that includes multiple siderostats, or a significant amount of background, can be related to the $V^2$ where only light from two siderostats is observed ($V_o ^2$), using the following expression:

\begin{equation}
V^2 = \frac{4 I_1 I_2 V_o ^2} {(\sum I_i)^2}.
\end{equation}

Here $I_1$ and $I_2$ are the fluxes from the two siderostats in a given baseline, while the sum in the denominator corresponds to the light from all siderostats and additional background observed in the same detector. Assuming that all siderostats have identical throughput and no additional background, one can calculate that in the case where 3 and 4 siderostats are included in the same spectrograph, their observed $V^2$'s are reduced by 4/9 and 1/4, respectively, relative to the case of a single baseline (i.e., 2 siderostats). Because SNR $\propto$ $NV^2$, we would expect a 9/16 reduction in the SNR when going from 3 to 4 siderostats in the same spectrograph. This would account for a significant portion of the increased scatter observed in the 6-way data presented in Figure \ref{HD187929_plots}, which always have 4 siderostats per spectrograph.

The other significant source of noise in the 6-way data is cross talk between the multiple baselines recorded in the same spectrograph. Due to the fact that the delay from the fast delay lines is modulated with stroke amplitudes in the range -4 to 4$\mu$m \citep{1998ApJ...496..550A}, non-linearities in the delay stroke amplitudes cause power from one baseline to spill into other baselines, affecting the fringe amplitudes and phases. Solutions to this problem include the recalibration of the strokes, an upgrade to new piezo electric actuators with longer stroke amplitudes, or to use the VISION beam combiner \citep{2016PASP..128e5004G}, which does not require the modulation of the delay. 

\begin{acknowledgements}

This material is based upon work supported by the National Aeronautics and Space Administration under Grant 18-XRP18$\_$2-0017 issued through the Exoplanets Research Program. The Navy Precision Optical Interferometer is a joint project of the Naval Research Laboratory and the U.S. Naval Observatory, and is funded by the Office of Naval Research and the Oceanographer of the Navy. This research has made use of the SIMBAD database and Vizier catalogue access tool, operated at CDS, Strasbourg, France. This work has made use of data from the European Space Agency (ESA) mission {\it Gaia} (\url{https://www.cosmos.esa.int/gaia}), processed by the {\it Gaia} Data Processing and Analysis Consortium (DPAC, \url{https://www.cosmos.esa.int/web/gaia/dpac/consortium}). Funding for the DPAC has been provided by national institutions, in particular the institutions participating in the {\it Gaia} Multilateral Agreement.

\end{acknowledgements}

\clearpage


\begin{deluxetable}{cccccccrc}
\tablecaption{Sample Star Properties.\label{general_properties}}
\tablehead{
 \colhead{}   & \colhead{}   & \colhead{}    & \colhead{Other} & \colhead{Spectral} & \colhead{$V$}   & \colhead{Parallax} & \colhead{}    & \colhead{}    \\
 \colhead{HD} & \colhead{HR} & \colhead{FK5} & \colhead{Name}  & \colhead{Type}     & \colhead{(mag)} & \colhead{(mas)}    & \colhead{Ref} & \colhead{[Fe/H]} \\ }
\startdata	
6186   & 294  & 36  & $\epsilon$ Psc & G9III     & 4.27 & 17.81 & 1 & -0.29	\\
10761  & 510  & 60  & o Psc          & G8III     & 4.26 & 12.53 & 1 & -0.03	\\
182640 & 7377 & 730 & $\delta$ Aql   & F1IV-V    & 3.36 & 64.41 & 2 & -0.04	\\
187929 & 7570 & 746 & $\eta$ Aql     & F6I+B9.8V & 3.73 & 2.61  & 3 & 0.13	\\
198001 & 7950 & 781 & $\epsilon$ Aqr & B9.5V     & 3.77 & 13.36 & 1 & -0.31	\\
210418 & 8450 & 834 & $\theta$ Peg   & A1V       & 3.52 & 36.77 & 3 & -0.38	\\
\enddata
\tablecomments{Spectral types are from SIMBAD, $V$ magnitudes are from \citet{Mermilliod}, parallaxes are from the following sources: 1. Gaia DR3 \citep{Gaia2022}; 2. \citet{2007AandA...474..653V}; 3. \citet{2018AandA...616A...1G}; and [Fe/H] is from \citet{2012AstL...38..331A}.}
\end{deluxetable}


\begin{deluxetable}{ccrlc}
\tablewidth{0pc}
\tabletypesize{\scriptsize}
\tablecaption{Observing Log.\label{observations}}
\tablehead{
 \colhead{Target}  & \colhead{Calibrator} & \colhead{Date} & \colhead{Baselines} & \colhead{$\#$} \\
 \colhead{HD}      & \colhead{HD}         & \colhead{(UT)} & \colhead{Used}     & \colhead{Data Points} \\ }
\startdata
6186   & 886    & 24 Aug 2021 & AC-AE, AC-AW, AC-E6, AC-W4, AC-W7, AE-AW, AE-W7, AW-E6, AW-W4, E6-W4        & 96  \\ 
       &        & 25 Aug 2021 & AC-AE, AC-AW, AC-E6, AC-W4, AC-W7, AE-AW, AE-W7, AW-E6, AW-W4, AW-W7, E6-W4 & 480 \\     
       &        &  5 Sep 2021 & AC-AE, AC-AW, AC-E6, AC-W4, AC-W7, AE-AW, AE-W7, AW-E6, AW-W4, AW-W7, E6-W4 & 630 \\
       &        &  8 Sep 2021 & AC-AE, AC-AW, AC-E6, AC-W4, AC-W7, AE-AW, AE-W7, AW-E6, AW-W4, AW-W7, E6-W4 & 960 \\       
       &        & 11 Sep 2021 & AC-AE, AC-AW, AC-E6, AC-W4, AC-W7, AE-AW, AE-W7, AW-E6, AW-W4, AW-W7, E6-W4 & 220 \\
       &        & 12 Sep 2021 & AC-AE, AC-AW, AC-E6, AC-W4, AC-W7, AE-AW, AE-W7, AW-E6, AW-W4, AW-W7, E6-W4 & 470 \\
       &        & 16 Sep 2021 & AC-AE, AC-AW, AC-E6, AC-W4, AC-W7, AE-AW, AE-W7, AW-E6, AW-W4, AW-W7, E6-W4 & 540 \\
10761  & 16582  &  5 Sep 2021 & AC-AE, AC-AW, AC-E6, AC-W4, AC-W7, AE-AW, AE-W7, AW-E6, AW-W4, AW-W7, E6-W4 & 950 \\
       &        &  7 Sep 2021 & AC-AE, AC-AW, AC-E6, AC-W4, AC-W7, AE-AW, AE-W7, AW-E6, AW-W4, AW-W7, E6-W4 & 360 \\
       &        & 11 Sep 2021 & AC-AE, AC-AW, AC-W4, AC-W7, AE-AW, AE-W7, AW-W4, AW-W7                      & 180 \\
       &        & 12 Sep 2021 & AC-AE, AC-AW, AC-E6, AC-W4, AC-W7, AE-AW, AE-W7, AW-E6, AW-W4, AW-W7, E6-W4 & 580 \\
       &        & 16 Sep 2021 & AC-AE, AC-AW, AC-E6, AC-W4, AC-W7, AE-AW, AE-W7, AW-E6, AW-W4, AW-W7, E6-W4 & 450 \\
182640 & 177756 & 13 Jun 2021 & AC-AE, AC-E6, AC-W4, E6-W4                                                  & 530 \\
       &        & 14 Jun 2021 & AC-AE, AC-E6, AC-W4, E6-W4                                                  & 490 \\
       &        & 25 Aug 2021 & AC-AE, AC-AW, AC-E6, AC-W4, AC-W7, AE-AW, AE-W7, AW-W4, AW-W7, E6-W4        & 555 \\ 
       &        & 27 Aug 2021 & AC-AE, AC-AW, AC-E6, AC-W4, AC-W7, AE-AW, AE-W7, AW-E6, AW-W4, AW-W7, E6-W4 & 1540 \\      
       &        & 28 Aug 2021 & AC-AE, AC-AW, AC-E6, AC-W4, AC-W7, AE-AW, AE-W7, AW-E6, AW-W4, AW-W7, E6-W4 & 660  \\
       &        &  5 Sep 2021 & AC-AE, AC-AW, AC-W4, AC-W7, AE-AW, AE-W7, AW-E6, AW-W4, AW-W7               & 558  \\
       &        &  6 Sep 2021 & AC-AE, AC-AW, AC-E6, AC-W4, AC-W7, AE-AW, AE-W7, AW-E6, AW-W4, AW-W7, E6-W4 & 2379 \\       
       &        &  8 Sep 2021 & AC-AE, AC-AW, AC-E6, AC-W4, AC-W7, AE-AW, AE-W7, AW-E6, AW-W4, AW-W7, E6-W4 & 850  \\       
       &        &  9 Sep 2021 & AC-AE, AC-AW, AC-E6, AC-W4, AC-W7, AE-AW, AE-W7, AW-E6, AW-W4, AW-W7, E6-W4 & 1477 \\
       &        & 12 Sep 2021 & AC-AE, AC-AW, AC-E6, AC-W4, AC-W7, AE-AW, AE-W7, AW-E6, AW-W4, AW-W7        & 599  \\       
       &        & 16 Sep 2021 & AC-AE, AC-AW, AC-W7, AE-AW, AE-W7, AW-W7                                    & 60   \\       
187929 & 184930 & 25 Aug 2021 & AC-AE, AC-AW, AC-E6, AC-W4, AC-W7, AE-AW, AE-W7, AW-E6, AW-W4, AW-W7, E6-W4 & 996  \\
198001 & 200761 &  5 Sep 2021 & AC-AE, AC-AW, AC-E6, AC-W4, AC-W7, AE-AW, AE-W7, AW-E6, AW-W4, AW-W7, E6-W4 & 1780 \\
       &        &  6 Sep 2021 & AC-AE, AC-AW, AC-E6, AC-W4, AC-W7, AE-AW, AE-W7, AW-E6, AW-W4, AW-W7, E6-W4 & 360  \\
       &        & 12 Sep 2021 & AC-AE, AC-AW, AC-E6, AC-W4, AC-W7, AE-AW, AE-W7, AW-E6, AW-W4, AW-W7, E6-W4 & 480 \\
       &        & 16 Sep 2021 & AC-AE, AC-AW, AC-E6, AC-W4, AC-W7, AE-AW, AE-W7, AW-E6, AW-W4, AW-W7, E6-W4 & 510 \\
210418 & 214923 & 24 Aug 2021 & AC-AE, AC-AW, AC-E6, AC-W4, AC-W7, AE-AW, AE-W7, AW-E6, AW-W4, E6-W4        & 655 \\
       &        & 27 Aug 2021 & AC-AE, AC-AW, AC-E6, AC-W4, AC-W7, AE-AW, AE-W7, AW-E6, AW-W4, AW-W7, E6-W4 & 840 \\      
       &        &  5 Sep 2021 & AC-AE, AC-AW, AC-E6, AC-W4, AC-W7, AE-AW, AE-W7, AW-E6, AW-W4, AW-W7, E6-W4 & 840 \\
       &        &  8 Sep 2021 & AC-AE, AC-AW, AC-W7, AE-AW, AE-W7, AW-W7                                    & 381 \\
       &        & 11 Sep 2021 & AC-AE, AC-AW, AC-W4, AC-W7, AE-AW, AE-W7, AW-W4, AW-W7                      & 90  \\      
       &        & 12 Sep 2021 & AC-AE, AC-AW, AC-E6, AC-W4, AC-W7, AE-AW, AE-W7, AW-E6, AW-W4, AW-W7, E6-W4 & 720 \\
\enddata
\tablecomments{See Table \ref{baselines} for the baseline lengths, and Figure \ref{array_config} for a representation of the configuration used.}
\end{deluxetable}

\clearpage


\begin{deluxetable}{cc}
\tablecaption{Baselines.\label{baselines}}
\tablehead{\colhead{Baseline}         & \colhead{Length (m)} }
\startdata	
\multicolumn{2}{c}{Spectrograph 1} \\	
\cline{1-2}
AC-AE & 18.9 \\	
AC-AW & 22.2 \\	
AC-W7 & 51.6 \\	
AE-AW & 44.1 \\	
AE-W7 & 64.4 \\	
AW-W7 & 29.5 \\
\cline{1-2}	
\multicolumn{2}{c}{Spectrograph 2} \\	
\cline{1-2}
AC-AW & 22.2 \\	
AC-E6 & 34.3 \\
AC-W4 & 8.8  \\	
AW-E6 & 53.3 \\	
AW-W4 & 14.0 \\	
E6-W4 & 42.5 \\	
\enddata
\end{deluxetable}


\begin{deluxetable}{cccccccccccccccc}
\tabletypesize{\scriptsize}
\tablecaption{Calibrator Stars' SED Inputs and Angular Diameters. \label{calibrators}}
\tablehead{
 \colhead{ }  & \colhead{Spec} & \colhead{$U$}   & \colhead{$B$}   & \colhead{$V$}   & \colhead{$R$}   & \colhead{$I$}   & \colhead{$J$}   & \colhead{$H$}   & \colhead{$K$}   & \colhead{$T_{\rm eff}$} & \colhead{log $g$}     & \colhead{}    & \colhead{}         & \colhead{}    & \colhead{$\theta_{\rm est}$} \\
 \colhead{HD} & \colhead{Type} & \colhead{(mag)} & \colhead{(mag)} & \colhead{(mag)} & \colhead{(mag)} & \colhead{(mag)} & \colhead{(mag)} & \colhead{(mag)} & \colhead{(mag)} & \colhead{(K)}           & \colhead{(cm s$^{-2}$)} & \colhead{Ref} & \colhead{$E(B-V)$} & \colhead{Ref} & \colhead{(mas)}   }
\startdata
886    & B2IV  & 1.75 & 2.61 & 2.83 & 2.88 & 3.06 & 3.50 & 3.64 & 3.77 & 21944 & 3.93 & 1 & 0.02 & 4 & 0.45$\pm$0.02 \\
16582  & B2IV  & 3.00 & 3.85 & 4.07 & 4.15 & 4.34 & 4.80 & 4.74 & 4.70 & 24118 & 4.19 & 2 & 0.00 & 5 & 0.23$\pm$0.01 \\
177756 & B8.5V & 3.07 & 3.34 & 3.43 & 3.44 & 3.52 & 3.52 & 3.48 & 3.56 & 11749 & 4.22 & 3 & 0.00 & 6 & 0.56$\pm$0.03 \\
184930 & B5III & 3.84 & 4.28 & 4.36 & 4.37 & 4.46 & 4.44 & 4.42 & 4.48 & 10471 & 3.72 & 3 & 0.07 & 7 & 0.45$\pm$0.02 \\
200761 & A1V   & 4.06 & 4.05 & 4.06 & 4.07 & 4.09 & 4.37 & 4.32 & 4.10 & 9550  & 4.01 & 3 & 0.01 & 8 & 0.50$\pm$0.03 \\
214923 & B8V   & 3.10 & 3.32 & 3.41 & 3.43 & 3.51 & 3.54 & 3.53 & 3.57 & 10965 & 3.75 & 3 & 0.01 & 9 & 0.60$\pm$0.03 \\
\enddata
\tablecomments{Spectral types are from SIMBAD; $UBV$ values are from \citet{Mermilliod}; $RI$ values are from \citet{2003AJ....125..984M}; $JHK$ values are from \citet{2003yCat.2246....0C}; $T_{\rm eff}$, log $g$, and $E(B-V)$ values are from the following sources: 1. \citet{2007astro.ph..3658P}; 2. \citet{2017MNRAS.471..770M}; 3. \citet{1999AandA...352..555A}; 4. \citet{2006MNRAS.371..703S}; 5. Gaia DR3 \citep{Gaia2022}; 6. \citet{1996AandAS..117..227A}; 7. \citet{2003AN....324..219W}; 8. \citet{1980BICDS..19...61N}; and 9. \citet{2009AandA...501..297Z}. $\theta_{\rm est}$ is the estimated angular diameter calculated using the method described in Section 2.}
\end{deluxetable}
	

\begin{deluxetable}{cccccccccccc}
\tablewidth{0pc}
\tablecaption{Interferometric Results. \label{inf_results}}

\tablehead{\colhead{Target} & \colhead{$\theta_{\rm UD}$} & \colhead{$T_{\rm eff}$} & \colhead{log $g$}        & \colhead{}    & \colhead{Initial}             & \colhead{$\theta_{\rm LD,initial}$} & \colhead{Final}               & \colhead{$\theta_{\rm LD,final}$} & \colhead{$\sigma_{\rm LD}$} & \colhead{Max SF} & \colhead{$\#$} \\ 
           \colhead{HD}     & \colhead{(mas)}             & \colhead{(K)}           & \colhead{(cm s$^{-2}$)}  & \colhead{Ref} & \colhead{$\mu_{\rm \lambda}$} & \colhead{(mas)}             & \colhead{$\mu_{\rm \lambda}$} & \colhead{(mas)}             & \colhead{($\%$)} & \colhead{(10$^6$ cycles s$^{\rm -1}$)}   & \colhead{pts}            }
\startdata
6186   & 1.813$\pm$0.025 & 4898 & 2.59 & 1 & 0.64 & 1.885$\pm$0.025 & 0.65 & 1.887$\pm$0.025 & 1.3  & 98.0 & 3396	\\
10761  & 1.583$\pm$0.018 & 5026 & 2.52 & 2 & 0.65 & 1.679$\pm$0.018 & 0.64 & 1.677$\pm$0.018 & 1.1  & 97.6 & 2520	\\
182640 & 1.163$\pm$0.016 & 7413 & 4.21 & 1 & 0.45 & 1.199$\pm$0.016 & 0.48 & 1.203$\pm$0.016 & 1.3  & 114.5 & 9698	\\
187929 & 1.713$\pm$0.055 & 5808 & 1.84 & 2 & 0.56 & 1.808$\pm$0.055 & 0.56 & 1.808$\pm$0.055 & 3.0  & 111.8 & 996	\\
198001 & 0.434$\pm$0.357 & 9120 & 3.55 & 1 & 0.42 & 0.504$\pm$0.357 & 0.39 & 0.503$\pm$0.357 & 71.0 & 95.6 & 3130	\\
210418 & 0.643$\pm$0.031 & 8511 & 4.02 & 1 & 0.45 & 0.689$\pm$0.031 & 0.43 & 0.688$\pm$0.031 & 4.5  & 97.4 & 3526	\\
\enddata 
\tablecomments{The initial $\mu_{\lambda}$ is based on the $T_{\rm eff}$ and log $g$ listed in the table, and the final $\mu_{\lambda}$ is based on the new $T_{\rm eff}$ determination. (See Section 3.2 for more details). The $T_{\rm eff}$ and log $g$ are from the following sources: 1. \citet{1999AandA...352..555A}; and 2. Gaia DR3 \citep{Gaia2022}. 
Max SF is the maximum spatial frequency for that star's diameter measurement. $\#$ pts is the number of data points in the angular diameter fit.}
\end{deluxetable}


\begin{deluxetable}{ccccccccc}
\tablewidth{0pc}
\tablecaption{Derived Stellar Parameters. \label{derived_results}}

\tablehead{\colhead{Target} & \colhead{Spectral}       & \colhead{$R$}         & \colhead{$\sigma_R$} & \colhead{$F_{\rm BOL}$}                      & \colhead{$A_{\rm V}$}           & \colhead{$T_{\rm eff}$} & \colhead{$\sigma_T$} & \colhead{$L$}        \\ 
           \colhead{HD}     & \colhead{Type}           & \colhead{($R_\odot$)} & \colhead{($\%$)}     & \colhead{(10$^{-6}$ erg s$^{-1}$ cm$^{-2}$)} & \colhead{(mag)}                 & \colhead{(K)}           & \colhead{($\%$)}        & \colhead{($L_\odot$)} }
\startdata
6186   & G9 III-IV & 11.39$\pm$0.19 & 1.7  & 0.648$\pm$0.003 & 0.08$\pm$0.01 & 4834$\pm$32    & 0.7  & 63.92$\pm$1.33	\\
10761  & G9 III-IV & 14.38$\pm$0.21 & 1.5  & 0.660$\pm$0.003 & 0.09$\pm$0.01 & 5152$\pm$28    & 0.6  & 131.50$\pm$2.75	\\
182640 & F0 IV-V   &  2.01$\pm$0.04 & 2.0  & 1.130$\pm$0.001 & 0.00$\pm$0.00 & 6958$\pm$46    & 0.7  & 8.52$\pm$0.26		\\
198001 & A0 V      &  4.05$\pm$2.87 & 71.0 & 0.920$\pm$0.004 & 0.00$\pm$0.00 & 10221$\pm$3627 & 35.5 & 161.18$\pm$9.04	\\
210418 & A3 III-IV &  2.01$\pm$0.11 & 5.5  & 0.961$\pm$0.005 & 0.00$\pm$0.00 & 8835$\pm$199   & 2.3  & 22.24$\pm$1.37	\\
\enddata
\tablecomments{The spectral types are those that provide the best SED fit as described in Section 3.2. The SED fits are also the source of $F_{\rm BOL}$ and $A_{\rm V}$, while the other parameters are derived as described also in Section 3.2. HD 187929/$\eta$ Aql is not included here due to its nature as a Cepheid variable, and the SED fit required to obtain these parameters is not usable.}
\end{deluxetable}


\clearpage

\clearpage

\clearpage

\clearpage

\end{document}